\documentclass[prl,aps,preprintnumbers,nofootinbib,twocolumn]{revtex4}
\usepackage{epsfig}
\usepackage{amsmath}
\usepackage{hyperref}

\def\as{\alpha_S}
\def\Lb{{\rm L}_\beta}
\def\Lr{{\rm L}_\rho}
\def\b{\beta}

\begin{document}

\preprint{CERN-PH-TH/2012-092,  TTK-12-13}

\renewcommand{\thefigure}{\arabic{figure}}

\title{Percent level precision physics at the Tevatron:\\ first genuine NNLO QCD corrections to $q\bar q \to t \bar t + X$}

\author{Peter B\"arnreuther}
\author{Micha\l{}  Czakon}
\affiliation{Institut f\"ur Theoretische Teilchenphysik und Kosmologie,
RWTH Aachen University, D-52056 Aachen, Germany}

\author{Alexander Mitov}
\affiliation{Theory Division, CERN, CH-1211 Geneva 23, Switzerland}

\date{\today}

\begin{abstract}
We compute the Next--to--Next--to--Leading Order (NNLO) QCD corrections to the partonic reaction that dominates top--pair production at the Tevatron. This is the first ever NNLO calculation of an observable with more than two colored partons, and/or massive fermions, at hadron colliders. Augmenting our fixed order calculation with soft-gluon resummation through Next--to--Next--to--Leading Logarithmic (NNLL) accuracy, we observe that the predicted total inclusive cross--section exhibits a very small perturbative uncertainty, estimated at $\pm2.7\%$. We expect that once all sub--dominant partonic reactions are accounted for, and work in this direction is ongoing, the perturbative theoretical uncertainty for this observable could drop below $\pm2\%$. Our calculation demonstrates the power of our computational approach and proves it can be successfully applied to all processes at hadron colliders for which high--precision analyses are needed. 
\end{abstract}
\maketitle

\section{Introduction}

Due to the number of special features it possesses, the top quark has become one of the pillars of the physics programs at the Tevatron and LHC. First, it has a uniquely large coupling to the Higgs sector of the Standard Model (SM). Second, the top quark is often the preferred decay mode of new massive objects, like heavy resonances, predicted in many models of new physics. Third, it decays before it hadronizes which makes it possible to study top quarks without encountering the non-perturbative effects that obscure the production of lighter quarks. Finally, and notably, one of the most significant current deviations from the SM is in the top quark forward--backward asymmetry $A_{FB}$ \cite{Aaltonen:2011kc}.

Motivated by these observations, in this work we make the first step towards a comprehensive increase of the predictive power of the Standard Model in the top sector. Specifically, we evaluate the Next--to--Next--to--Leading Order (NNLO) QCD corrections to the total inclusive top-pair cross-section for the reaction dominating at the Tevatron, $q\bar q\to t\bar t +X$. Our Tevatron prediction, based on this first ever NNLO calculation for a hadron collider process involving more than two colored partons and/or massive fermions, is almost twice as precise as the currently available experimental measurements. We believe that our results are a strong motivation for further experimental improvements in top physics.

We expect to extend our work with fully differential results for top--pair production at the Tevatron and the LHC, as well as production of dijets, $W+{\rm jet}$, Higgs$ + {\rm jet}$, etc. All these processes will be instrumental in the ongoing and future searches for new physics and for assessing the workings of the Standard Model at the finest level.

\section{Top production at hadron colliders}

The total top-pair production cross-section reads:
\begin{equation}
\sigma_{\rm tot} = \sum_{i,j} \int_0^{\beta_{\rm max}}d\beta\, \Phi_{ij}(\beta,\mu^2)\, \hat\sigma_{ij}(\beta,m^2,\mu^2)  \, ,
\label{eq:sigmatot}
\end{equation}
where $i,j$ run over all possible initial sate partons, $\Phi_{ij}$ is the partonic flux which is a convolution of the densities of partons $i,j$ and includes a Jacobian factor. The dimensionless variable $\beta=\sqrt{1-\rho}$, with $\rho = 4m^2/s$, is the relative velocity of the final state top quarks having pole mass $m$ and produced at partonic c.m. energy $\sqrt{s}$; $\mu$ stands for both the renormalization ($\mu_R$) and factorization scales ($\mu_F$). 

For $\mu_F=\mu_R=\mu$ the partonic cross-section reads:
\begin{eqnarray}
&&\hat\sigma_{ij}\left(\beta,m^2,\mu^2\right) = {\as^2\over m^2}\Bigg\{  \sigma^{(0)}_{ij} + \as \left[ \sigma^{(1)}_{ij} + L\, \sigma^{(1,1)}_{ij} \right] + \nonumber\\
&& \as^2\left[ \sigma^{(2)}_{ij} + L\, \sigma^{(2,1)}_{ij} + L^2 \sigma^{(2,2)}_{ij} \right] + {\cal O}(\as^3) \Bigg\} \, ,
\label{eq:sigmapart}
\end{eqnarray}
where $L = \ln\left(\mu^2/m^2\right)$ and $\as$ is the ${\overline {\rm MS}}$ coupling renormalized with $N_L=5$ active flavors at scale $\mu^2$. The functions $\sigma^{(n(,m))}_{ij}$ depend only on $\beta$.

The dependence on $\mu_R\neq \mu_F$ can be trivially restored in Eq.~(\ref{eq:sigmapart}). The LO result reads
$\sigma^{(0)}_{q\bar q} = \pi\beta\rho(2+\rho)/27$. The NLO results are known exactly \cite{Nason:1987xz}. The scale controlling functions $\sigma^{(2,1)}_{ij}$ and $\sigma^{(2,2)}_{ij}$ can be easily computed from the NLO results $\sigma^{(1)}_{ij}$, and can be found in \cite{Langenfeld:2009wd}.

In this work, for the first time, complete results for the function $\sigma^{(2)}_{q\bar q}$ are derived. Work towards the calculation of the remaining reactions $ij=gg,gq,qq'$ is underway. We recall that the only information about $\sigma^{(2)}_{q\bar q}$ and $\sigma^{(2)}_{gg}$ available so far was from their $\beta \to 0$ limits \cite{Beneke:2009ye}.

\section{Parton level results}

In this paper we calculate the NNLO correction $\sigma^{(2)}_{q\bar q}$ to the partonic reaction $q \bar q \to t \bar t + X$.  The result reads:
\begin{eqnarray}
\sigma^{(2)}_{q\bar q}(\beta) = F_0(\beta) +F_1(\beta) N_L +F_2(\beta) N_L^2  \, ,
\label{eq:sigma2qq}
\end{eqnarray}
The dependence on the number of light flavors $N_L$ in Eq.~(\ref{eq:sigma2qq}) is explicit. The function $F_2$ is derived exactly:
\begin{eqnarray}
F_2 = {\sigma^{(0)}_{q\bar q} \over 108 \pi^2}\left( 25-3 \pi^2+30 \ln\left({\rho\over 4}\right)+9 \ln^2\left({\rho\over 4}\right) \right) \, .
\label{eq:F2}
\end{eqnarray}

The functions $F_i \equiv F^{(\beta)}_i + F^{(\rm fit)}_i, i=0,1,$ read:
\begin{eqnarray}
F^{(\beta)}_1 &=& \sigma^{(0)}_{q\bar q}\left[ (0.0116822 - 0.0277778 \Lb)/\beta \right.\nonumber\\ 
&&\left.  + 0.353207 \Lb - 0.575239 \Lb^2 +  0.240169 \Lb^3\right] \, , \label{eq:F1beta0} \\ 
F^{(\beta)}_0 &=&  \sigma^{(0)}_{q\bar q}\left[0.0228463/\beta^2 +(-0.0333905+0.342203 \Lb \right.\nonumber\\
&& -0.888889 \Lb^2)/\beta+1.58109 \Lb+6.62693 \Lb^2 \nonumber\\
&&\left. -9.53153 \Lb^3+5.76405 \Lb^4 \right]  \, , \label{eq:F0beta0}\\ 
F^{(\rm fit)}_1 &=& (0.90756 \b-6.75556 \b^2+9.18183 \b^3) \rho \nonumber\\
&&+(-0.99749 \b+27.7454 \b^2-12.9055 \b^3) \rho^2\nonumber\\
&&+(-0.0077383 \b-4.49375 \b^2+3.86854 \b^3) \rho^3\nonumber\\
&&-0.380386 \b^4 \rho^4+\Lr (1.3894 \rho+6.13693 \rho^2\nonumber\\
&&+8.78276 \rho^3-0.0504095 \rho^4) +\Lr^2\, 0.137881 \rho \, , \label{eq:F1fit}\\ 
F^{(\rm fit)}_0 &=&  (-2.32235 \b+44.3522 \b^2-24.6747 \b^3) \rho\nonumber\\
&&+(2.92101 \b+224.311 \b^2+21.5307 \b^3) \rho^2\nonumber\\
&&+(2.05531 \b+945.506 \b^2+36.1101 \b^3\nonumber\\
&&-176.632 \b^4) \rho^3+7.68918 \b^4 \rho^4+\Lr (3.11129 \rho\nonumber\\
&&+100.125 \rho^2+563.1 \rho^3+568.023 \rho^4) \, , \label{eq:F0fit}
\end{eqnarray}
where $\Lr \equiv \ln(\rho)$ and $\Lb \equiv \ln(\beta)$. The functions $F^{(\beta)}_{1,0}$ constitute the threshold approximation to $\sigma^{(2)}_{q\bar q}$ \cite{Beneke:2009ye} multiplied by the full Born cross-section $\sigma^{(0)}_{q\bar q}$ and with the constant $C^{(2)}_{q\bar q}$ (as introduced in \cite{Beneke:2009ye}) set to zero.

\begin{figure}[t]
  \centering
     \hspace{0mm} 
   \includegraphics[width=0.51\textwidth]{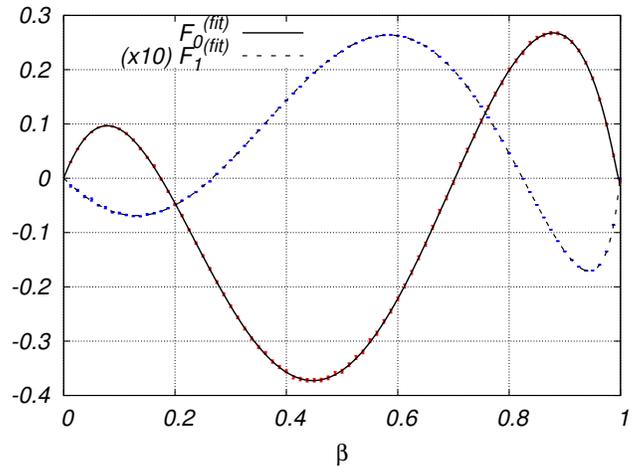} 
   \caption{ \label{fig:fit} The functions $F^{(\rm fit)}_{0}$ and $10\times F^{(\rm fit)}_{1}$ (rescaled for improved visibility) as: a) discrete sets of calculated values, with errors, on the grid of 80 points (red and blue error bars) and b) the analytical fits given in Eqns.~(\ref{eq:F0fit},\ref{eq:F1fit}) (black lines).}
\end{figure}
 
The functions $F_{1,0}$ are computed numerically on a grid of 80 points in the interval $\beta \in (0,1)$. The functions $F^{(\rm fit)}_i$ are then derived as a fit to the difference $F_i-F^{(\beta)}_{i}$.

On Fig.~\ref{fig:fit} we present the fitted functions $F^{(\rm fit)}_{i}$ together with the calculated values for $F_i-F^{(\beta)}_{i}$ in all 80 points, including the estimated numerical errors for each evaluation point. We note that the precision of our result is not limited by the quality of the fit, but rather by the precision of the numerical evaluation of $F_{1,0}$. The absolute error on $\sigma^{(2)}_{q\bar q}$, for $N_L = 5$, is bounded by $3.6\times 10^{-3}$. At the Tevatron this translates into a relative error on the cross-section of $3\times 10^{-4}$, which is negligible.

The fits become unreliable very close to the high-energy limit $\beta\to 1$, i.e. beyond the last computed point $\beta_{80}=0.999$. While this loss of precision is completely immaterial for top production ($\beta_{80} \equiv 0.999$ is approximately the value of $\beta_{\rm max}$ for the LHC at 8 TeV), it might be an impediment for the description of lighter quark production, like bottom. The reason is that for very light quarks the partonic flux $\Phi$ becomes strongly peaked towards $\beta \approx 1$ which makes the hadronic cross-section sensitive to the behavior of the fits in this limit.

From Eqn.~(\ref{eq:sigma2qq}) we can extract an approximate value for the two--loop constant term $C^{(2)}_{q\bar q}$, as defined in \cite{Beneke:2009ye}, which translates into the hard matching constant $H^{(2)}_{q\bar q}$ as defined in Ref.~\cite{Cacciari:2011hy}. We get the following values:
\begin{eqnarray}
C^{(2)}_{q\bar q} &=& 1195.82 -44.1841 N_L-4.28168 N_L^2 \, , \\
H^{(2)}_{q\bar q} &=& 84.8139 ~ ({\rm for}\, N_L=5) \, .
\label{eq:C2H2}
\end{eqnarray}

Following the findings of Refs.~\cite{Hagiwara:2008df}, we caution about the accuracy of the extraction of $C^{(2)}_{q\bar q}$ (and therefore $H^{(2)}_{q\bar q}$). Assuming a polynomial in $\beta$ form for the fits $F^{(\rm fit)}_{1,0}$, we can extract $C^{(2)}_{q\bar q}$ with a precision better than $10\%$ (which implies $H^{(2)}_{q\bar q}\in (80,90))$. This uncertainty has sub-permil numerical effect for top production at the Tevatron. On the other side, we note that if we allow into the fits terms containing powers of $\ln(\beta)$, then $C^{(2)}_{q\bar q}$ cannot be extracted with any reasonable precision (the reason being the finite size of the grid). At any rate, the overall smooth behavior of the fits suggests that our extraction is reliable. 

\begin{figure}[t]
  \centering
     \hspace{0mm} 
   \includegraphics[width=0.51\textwidth]{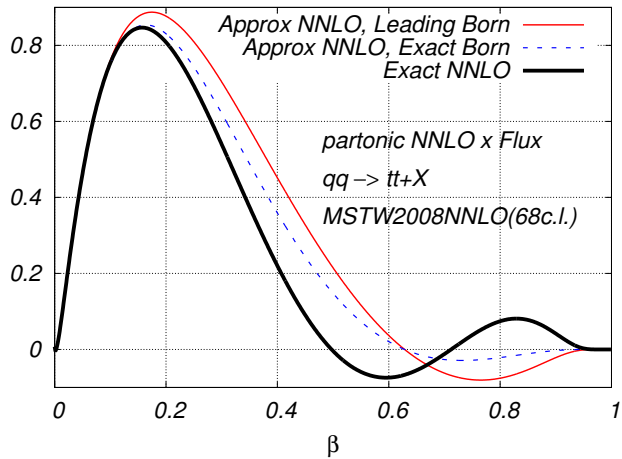} 
   \caption{ \label{fig:with-fluxes} Partonic cross-section times the partonic flux for Tevatron (same as the integrand of Eq.~(\ref{eq:sigmatot})) for three different partonic cross-section:  a) exact NNLO (black thick line), b) ${\rm NNLO}_{\rm approx}$ with exact Born $\sigma^{(0)}_{q\bar q}$ (blue dashed line) and c) ${\rm NNLO}_{\rm approx}$ with leading Born $\sigma^{(0)}_{q\bar q}|_{\beta\to 0}$ (red thin line). }
\end{figure}

It is interesting to compare the exact partonic cross-section Eq.~(\ref{eq:sigma2qq}) with the approximately known one \cite{Beneke:2009ye}. To that end on Fig.~\ref{fig:with-fluxes} we plot the partonic cross-section multiplied by the partonic flux for Tevatron (i.e. the integrand of Eq.~(\ref{eq:sigmatot})) for the following three cases: a) exact NNLO (\ref{eq:sigma2qq}), b) approximate NNLO defined by setting $F_2$ and $F^{(\rm fit)}_i$ in (\ref{eq:sigma2qq}) to zero and c) as in b) with the additional replacement $\sigma^{(0)}_{q\bar q} \longrightarrow \sigma^{(0)}_{q\bar q}|_{\beta\to 0} = \pi\beta/9$ in Eqns.~(\ref{eq:F1beta0},\ref{eq:F0beta0}). We observe that the approximate expression strongly depends on subleading power effects and is not a very good approximation for the exact result. Upon integration, these differences get reduced due to accidental cancellation in the intermediate $\beta$ region where the approximate results are smaller/larger than the exact one.

Before closing this section we briefly explain our calculational approach. The two-loop virtual corrections are taken from \cite{Czakon:2008zk}, utilizing the analytical form for the poles \cite{Ferroglia:2009ii}. The one-loop squared amplitude is computed in \cite{Korner:2008bn}. The real-virtual corrections are derived by integrating the one--loop amplitude with a counter-term that regulates it in all singular limits \cite{Bern:1999ry}. The finite part of the one-loop amplitude is computed with a code used in the calculation of $pp\to t\bar t + {\rm jet}$ at NLO \cite{Dittmaier:2007wz}. The double real corrections are computed in \cite{Czakon:2010td}. As in Ref.~\cite{Czakon:2010td}, we do not include the contribution from the reaction $q\bar q \to t\bar t q\bar q$ where the final state light pair has the same flavor as the initial state one. We expect the contribution from this reaction to be negligible. We consistently modify the collinear subtraction to account for this missing contribution. Details will be presented elsewhere.

\section{Numerical predictions: the Tevatron}\label{sec:Tevatron}

The NNLO results computed in this paper make it possible to predict the total top-pair cross-section at the Tevatron with high precision. Implementing the new NNLO results in the program {\tt Top++} \cite{Czakon:2011xx} (with $m_t=173.3~{\rm GeV}$ and MSTW2008nnlo68cl pdf set~\cite{Martin:2009iq} throughout) and adopting the scale and pdf variation procedures of Ref.~\cite{Cacciari:2011hy} we get (in {\rm pb}):
\begin{eqnarray}
\sigma_{\rm tot}^{\mathrm{NNLO}} = 7.005^{~+0.202\, (2.9\%)}_{~- 0.310\,(4.4\%)}~[{\rm scales}] 
^{~+ 0.170\,(2.4\%)}_{~- 0.122\, (1.7\%)}~[{\rm pdf}]\, .~~
\label{eq:best-result-FO}
\end{eqnarray}

The fixed order NNLO result $\sigma_{\rm tot}^{\mathrm{NNLO}}$ includes the complete NNLO $q\bar q$ contribution (\ref{eq:sigma2qq}) and the approximate NNLO result for the $gg$ reaction as implemented in \cite{Cacciari:2011hy}\cite{Czakon:2011xx}. We have set the unknown constant $C^{(2)}_{gg}=0$. We have verified that the sensitivity to the exact value of  $C^{(2)}_{gg}$ is around $\pm 0.5\%$ when $C^{(2)}_{gg}$ is varied in the range $\pm 1137$, i.e. the value of the constant ${\overline C}^{(2,0)}_{gg}$  \cite{Cacciari:2011hy}. 

Our best prediction $\sigma_{\rm tot}^{\mathrm{res}} \equiv \sigma_{\rm tot}^{\mathrm{NNLO}+\mathrm{NNLL}}$ is derived with full NNLL resumaton \cite{Beneke:2009rj} matched to the exact NNLO result for the $q\bar q$ reaction, including $H^{(2)}_{q\bar q}$ (\ref{eq:C2H2}):
\begin{eqnarray}
\sigma_{\rm tot}^{\mathrm{res}} = 7.067^{~+ 0.143\, (2.0\%)}_{~- 0.232\,(3.3\%)}~[{\rm scales}]^{~+ 0.186 \,(2.6\%)}_{~- 0.122\, (1.7\%)} ~[{\rm pdf}]\, , ~~
\label{eq:best-result-res}
\end{eqnarray}
while the contribution from the $gg$ reaction to Eq.~(\ref{eq:best-result-res}) is implemented as ${\rm NNLO}_{\rm approx}$+NNLL \cite{Cacciari:2011hy}\cite{Czakon:2011xx}. We take $A=2$ for the value of the constant $A$ \cite{Bonciani:1998vc} that controls power suppressed effects.

We find a $0.4\%$ sensitivity of $\sigma_{\rm tot}^{\mathrm{NNLO}+\mathrm{NNLL}}$ to the value of the constant $A$. To be conservative in our error estimate, we exclude the scale dependent term at the level of the unknown two-loop constant in the $gg$ reaction, see \cite{Cacciari:2011hy}\cite{Czakon:2011xx}. Their inclusion lowers the scale uncertainty from $\pm 2.7\%$ to $\pm 1.7\%$. On the other side, including these scale dependent terms brings about  ${\cal O}(1\%)$ sensitivity to the value of the unknown hard matching coefficients $H^{(2)}_{gg,{\bf 1}}, H^{(2)}_{gg,{\bf 8}}$ which offsets the reduction in scale variation. 

We conclude that the error estimate of our best result Eq.~(\ref{eq:best-result-res}) takes into account all dominant sources of theoretical uncertainty, and that the missing NNLO contributions from other reactions will affect the above results at the percent level, i.e. they are accounted for by our theoretical uncertainty.

\begin{figure}[t]
  \centering
     \hspace{0mm} 
   \includegraphics[width=0.49\textwidth]{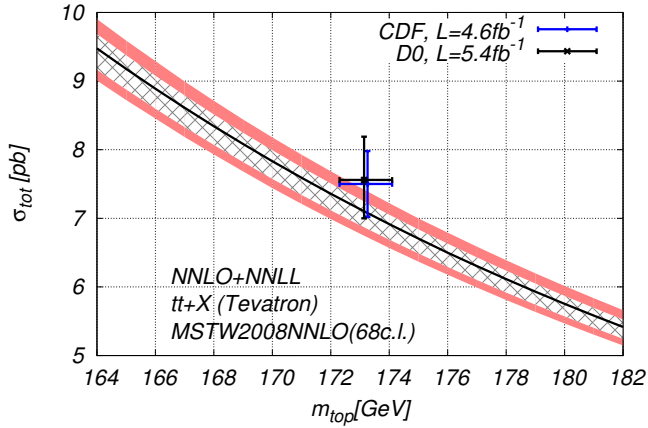} 
   \caption{ \label{fig:Tev-mass-range} Dependence of the NNLO+NNLL cross-section $\sigma_{\rm tot}^{\mathrm{NNLO}+\mathrm{NNLL}}$, Eq.~(\ref{eq:best-result-res}), on the pole mass of the top quark: scale variation (white band); scale + pdf variation (red band).}
\end{figure}

On Fig.~\ref{fig:Tev-mass-range} we present the dependence of our best prediction $\sigma_{\rm tot}^{\mathrm{NNLO}+\mathrm{NNLL}}$ on the value of the top mass. It includes scale and pdf variation. We find very good agreement with the latest measurements from the Tevatron \cite{Abazov:2011cq} and note that the total theoretical uncertainty is only about one-half of the total experimental one.

\section{Conclusions and Outlook}

In this work we calculate the genuine NNLO corrections to the total inclusive cross-section for $q\bar q\to t\bar t+X$. After extracting the two-loop hard matching constant from our result, we augment the NNLO evaluation with soft gluon resummation with full NNLL accuracy. As anticipated, our NNLO+NNLL numerical prediction for the Tevatron has substantially improved precision in comparison with NLO+NNLL or approximate NNLO results. Most importantly, the accuracy of our theoretical prediction exceeds the accuracy of the best currently available measurements from the Tevatron. We are confident that our results will provide new insight to the forthcoming Tevatron analyses at full dataset, and will help scrutinize the Standard Model to a new level. 

The very high precision of our result will allow critical comparisons between different pdf sets as well as extraction of the top quark  mass with improved precision. It is also a step in the derivation of the dominant missing SM corrections to $A_{FB}$, whose calculation through order ${\cal O}(\as^4)$ will be the subject of a forthcoming  publication.

\begin{figure}[t]
  \centering
     \hspace{0mm} 
   \includegraphics[width=0.49\textwidth]{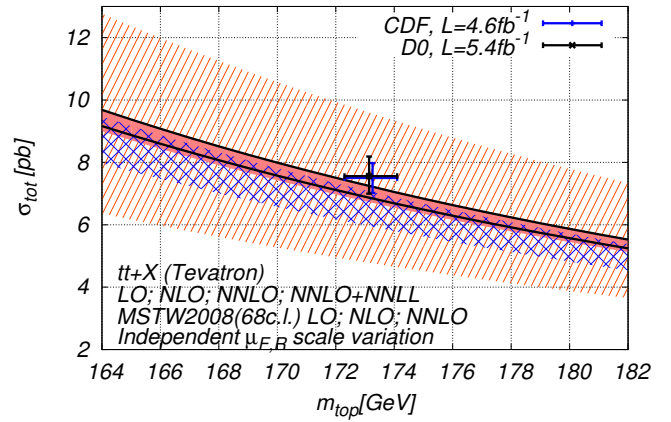} 
   \caption{ \label{fig:LO-NLO-NNLO-res} Scale variation at LO (orange), NLO (blue), NNLO (red) and resumed NNLO+NNLL (black lines).}
\end{figure}

In a  broader context, given the small number of observables known at NNLO, it is interesting to address the question of the convergence of the perturbative series for this observable. In Fig.~(\ref{fig:LO-NLO-NNLO-res}) we plot the scale variations of the LO, NLO, NNLO and NNLO+NNLL approximations as functions of the top mass. Each approximation is calculated with a pdf of corresponding accuracy. 
We observe a significant and consistent decrease in the scale dependence with each successive approximation. The overlap between the scale bands of the successive approximations also indicates that our scale variation procedure performs consistently well at each perturbative order.

\begin{acknowledgments}
We thank S.~Dittmaier for kindly providing us with his code for the evaluation of the one-loop virtual corrections in $q\bar q \to t\bar t g$ \cite{Dittmaier:2007wz}, and Z.~Merebashvili for clarifications regarding Ref.~\cite{Korner:2008bn}. The work of M.C. was supported by the Heisenberg and by the Gottfried Wilhelm Leibniz programmes of the Deutsche Forschungsgemeinschaft, and by the DFG Sonderforschungsbereich/Transregio 9 ÒComputergest\"utzte Theoretische TeilchenphysikÓ.
\end{acknowledgments}

\end{document}